\documentclass[preprint,showpacs,preprintnumbers,amsmath,amssymb]{revtex4}

\usepackage{graphicx}
\usepackage{dcolumn}
\usepackage{bm}
\usepackage{color}
\usepackage{ulem}

\begin{document}


\title{Neutron lifetime measurements using gravitationally trapped ultracold
 neutrons}

\author{A.~P.~Serebrov,$^{1,}$\footnote{serebrov@pnpi.spb.ru}
V.~E.~Varlamov,$^1$ A.~G.~Kharitonov,$^1$ A.~K.~Fomin,$^1$
Yu.~N.~Pokotilovski,$^2$ P.~Geltenbort,$^3$ I.~A.~Krasnoschekova,$^1$
 M.~S.~Lasakov,$^1$
R.~R.~Taldaev,$^1$ A.~V.~Vassiljev,$^1$ and O.~M.~Zherebtsov$^1$}

\affiliation{
$^{1}$Petersburg Nuclear Physics Institute, Russian Academy of Sciences,
 188300 Gatchina, Leningrad District, Russia\\
$^{2}$Joint Institute for Nuclear Research, 141980 Dubna, Moscow Region,
 Russia\\
$^{3}$Institut Max von Laue Paul Langevin, BP 156, F-38042 Grenoble Cedex 9,
 France}


\begin{abstract}
Our experiment using gravitationally trapped ultracold neutrons
(UCN) to measure the neutron lifetime is reviewed. Ultracold neutrons were trapped in
a material bottle covered with perfluoropolyether.  The neutron
lifetime was deduced from comparison of UCN losses in the traps with
different surface-to-volume ratios. The precise
value of the neutron lifetime is of fundamental importance to
particle physics and cosmology. In this experiment, the UCN storage
time is brought closer to the neutron lifetime than in any experiments before:
the probability of UCN losses from the trap was only 1$\%$ of that
for neutron $\beta$ decay. The neutron lifetime obtained,
$878.5\pm 0.7_{stat}\pm 0.3_{sys}$   s, is the most accurate experimental
 measurement to date.
\end{abstract}

\maketitle

\section{Introduction}

Precision measurements of the neutron lifetime are 
important for elementary particle physics and cosmology. The decay
of a free neutron into a proton, an electron, and an antineutrino
is determined by the weak interaction comprising the transition of
a \textit{d} quark into a \textit{u} quark. In the Standard Model of elementary
particles, the quark mixing is described by the Cabibbo-Kobayashi-Maskawa
(CKM) matrix which must be unitary. For
instance, for the first row we have
\begin{equation}
\label{eq1}
|V_{ud} |^2  + |V_{us} |^2  + |V_{ub} |^2  = {\rm  }1
\end{equation}
where $V_{ud} ,V_{us}$, and $V_{ub}$ are the matrix elements related to
 the mixing of a \textit{u} quark with a \textit{d}, \textit{s}, or \textit{b} quark
respectively. The values of the individual matrix elements are
determined by the weak decays of the respective quarks. In
particular, the matrix element   $V_{ud}$ can be determined from
a nuclear $\beta$ decay and neutron $\beta$ decay. The
extraction of $V_{ud}$ from neutron $\beta$ decay data is
attractive because of the theoretical simplicity in describing
neutron decay. The experimental procedure itself requires
 precise measurements of the neutron
lifetime  $\tau _n$ and the $\beta$ decay asymmetry $A_0$. The
neutron half-life $t$ is given by the following equation
\cite{Abe04}:
\begin{equation}
\label{eq2} ft(1 + \delta '_R ) = \frac{K}{{|V_{ud} |^2 G_F ^2 (1
+ 3\lambda ^2 )(1 + \Delta _R )}}
\end{equation}
where  $f=1.6886$ is the phase space factor, $\delta
'_R=1.466\times10^{-2}$ is a model-independent external radiative
correction \cite{Wil82,Mar06},  $\Delta _R=2.40\times10^{-2}$ is a
model-dependent internal radiative correction \cite{Tow92,Tow03},
$\lambda = G_A /G_V $ - is the ratio of the axial-vector weak coupling
constant to the vector weak coupling constant,  $G_F$  is the Fermi weak coupling constant determined
from the  $ \mu $ decay, and  $K$ is a combination of the known
fundamental constants. The relative uncertainties in the
electroweak radiative corrections are of the order of a few percent.
The general formula for $|V_{ud} |^2$  as a function of
$\tau _n$ and $\lambda$ takes the form \cite{Mar06}
\begin{equation}
\label{eq3} \left| {V_{ud} } \right|^2  =  \frac{{4908.7 \pm
1.9~s}}{{\tau _n (1 + 3\lambda ^2 )}}
\end{equation}
where the accuracy in calculation of the radiative corrections
has been incorporated. Thus, the required relative accuracy of
the neutron lifetime  $\tau _n$ measurement must be higher than
$10^{-3}$, whereas the relative accuracy of the $\lambda$ measurement must
be higher than $0.5\times10^{-3}$. The parameter  $\lambda$ can be
obtained from the measurements of the asymmetry $A_0$ of the neutron
$\beta$ decay:
\begin{equation}
\label{eq4}
A_0  =  - 2\frac{{\lambda (\lambda  + 1)}}{{1 + 3\lambda ^2 }}
\end{equation}
Since  $\Delta \lambda /\lambda  = 0.25\Delta A_0 /A_0$, the
relative accuracy of the asymmetry measurement must be higher than
$2\times10^{-3}$.

Precise measurements of the neutron lifetime are also
important input parameters in the models of the early stages of
the formation of the Universe.

The observed quantities in the Big Bang model are the initial
abundances of deuterium and ${}^4He$. These quantities depend on the ratio of
the number of baryons to the number of photons in the initial
nucleosynthesis stage and on the neutron lifetime  $\tau _n$. For
instance, at a fixed value of baryon asymmetry  $\eta _{10}$, a
variation in the neutron lifetime by 1$\%$ changes the value of
the initial abundance of ${}^4He$ by 0.75$\%$. The relative
accuracy of the measurement of ${}^4He$ abundance is ±0.61$\%$
\cite{Mat05}. Similarly, a variation in the neutron lifetime by
1$\%$ changes  $\eta _{10}$ by 17$\%$, the value of
$\eta _{10}$ is currently estimated to the precision of ±3.3$\%$
\cite{Mat05}. Thus, to verify the nucleosynthesis model in the Big
Bang, the accuracy of the neutron lifetime measurement must be
higher than 1$\%$.

The results presented in this paper have already published briefly
in Ref. \cite{Ser05}. Here we present a more detailed account of the experiment.

\section{Experimental setup}

\begin{figure}
\includegraphics[width=3.5in]{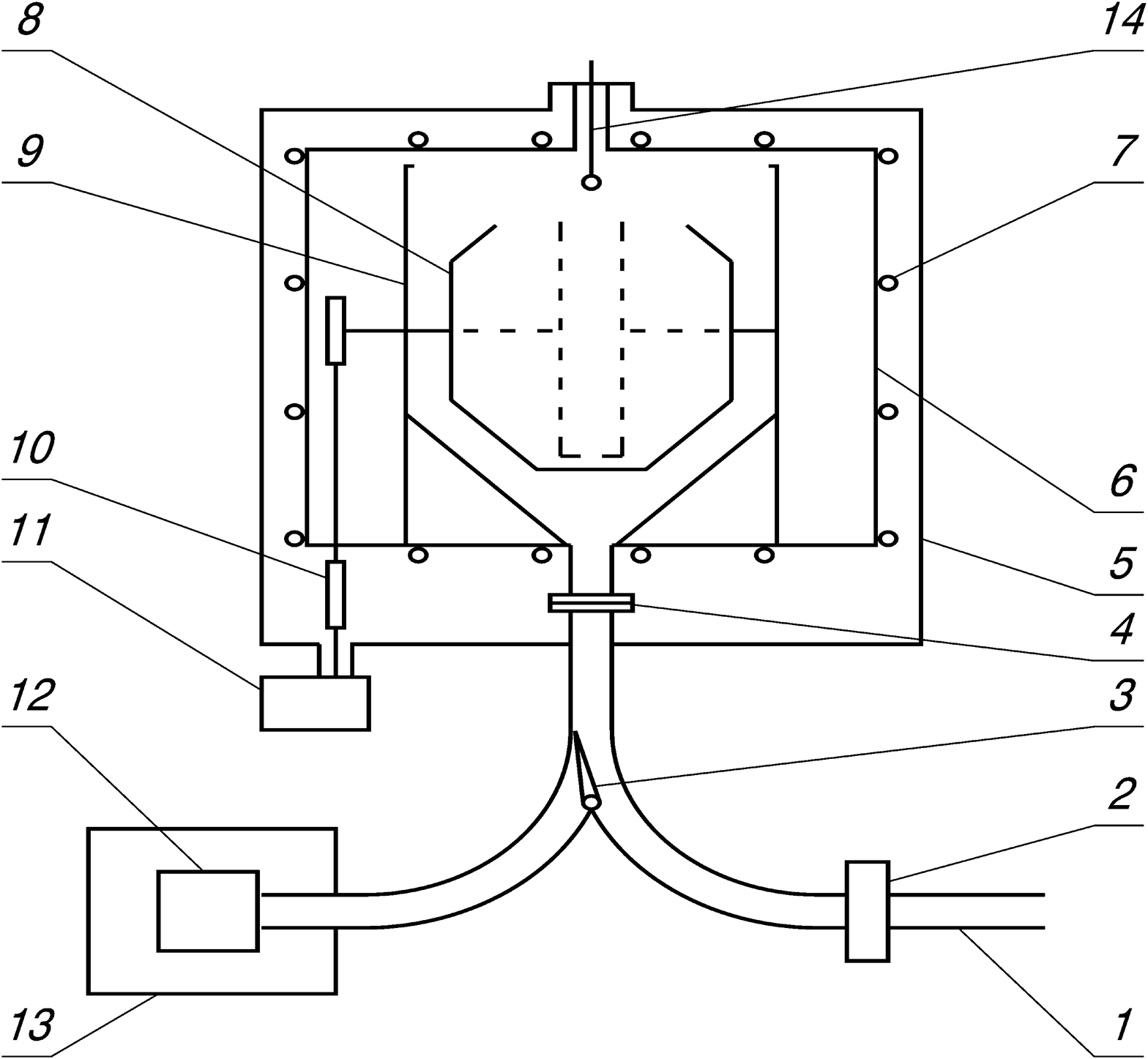}
\caption{\label{fig1} Schematic of the gravitational UCN storage
system: 1--input neutron guide for UCN; 2--inlet valve; 3--selector
valve (shown in the position in which the trap is being
filled with neutrons); 4--foil unit; 5--vacuum volume; 6--separate
vacuum volume of the cryostat; 7--cooling system for the
thermal shields; 8--UCN storage trap (the dashed lines depict a
narrow cylindrical trap); 9--cryostat; 10--trap rotation drive;
11--step motor; 12--UCN detector; 13--detector shield, and
14--vaporizer.}
\end{figure}

The experimental facility was a joint project of the
B.~P.~Konstantinov Petersburg Nuclear Physics Institute (PNPI),
Gatchina and the Joint Institute for Nuclear Research (JINR),
Dubna. The experimental setup was used for the first time at the universal source of
cold and ultracold neutrons from the water-moderated water-cooled
reactor WWR-M in Gatchina (Russia). The cooling of the
facility to 10--15~K was made with a refrigerator. Later, the equipment
was modified to the cryostat scheme in order to be able to run it at 
the high-flux ILL reactor in Grenoble (France). Figure~\ref{fig1} 
shows the modified version of the experimental setup.

The experiment comprises a gravitational trap for UCN, which can
also serve as a differential gravitational spectrometer. Hence, a
distinctive feature of this experimental setup is its possibility
of measuring the UCN energy spectrum after the neutron has been
stored in the trap.

The UCN storage trap (8) is placed inside the vacuum volume of the
cryostat (9). The trap has a window and can be rotated around the
horizontal axis in such a way that the UCN are held by the gravitational
field in the trap when the window is in its uppermost position.

The ultracold neutrons enter the trap through the neutron guide (1),
after passing the inlet valve (2) and the selector valve (3).
Filling of the trap by the ultracold gas is done with the trap
window in the ``down'' position. After filling, the trap is
rotated $180^\circ$ so that the window is in the ``up'' position.

The vacuum system comprises two separate vacuum volumes: the
``high-vacuum'' and the ``isolating'' volume. The pressure in the
high-vacuum volume of the cryostat is $5\times10^{-6}$
mbar. At this pressure, the residual gas affects only slightly
the storage time (by about 0.4 s; see Sec.~\ref{residual gas section})
of the UCN in the trap. The trap is cooled through the heat exchange
between the trap and the cryostat's reservoir. To improve the heat
exchange, gaseous helium was blown through the vacuum volume of
the cryostat and later removed before measuring the neutron lifetime.

The position (height) of the trap window, with respect to the bottom
of the trap, determines the maximum energy of the UCN that can be
stored in the trap. The different values of the window's height
correspond to the different values of the cutoff energy in the UCN
spectrum; this rotating trap is thus a gravitational spectrometer.
The spectral dependence of the neutron
storage time can be measured by a series of measurements whereby one
varies the trap window height. The trap was kept in
each position for 100--150 s to register the UCN
in that energy range. We measured the spectrum
of the trapped UCN according to this procedure.

The neutron lifetime was measured by the so-called size extrapolation
method. To this end, two UCN traps with different dimensions were
employed. The first trap was quasi-spherical, consisting of a
horizontal cylinder of 26 cm in length and 84 cm in diameter that was
``crowned'' by two 22-cm-high truncated cones with a smaller diameter
of 42 cm. The second trap was
cylindrical, with the length of 14 cm and 76 cm in diameter. The
frequency of neutron collisions with the walls of the second trap
was approximately 2.5 times higher than that in the first trap. In
Fig.~\ref{fig1}, the narrow cylindrical trap is depicted by dashed
lines.

\begin{figure}
\includegraphics[width=3.5in]{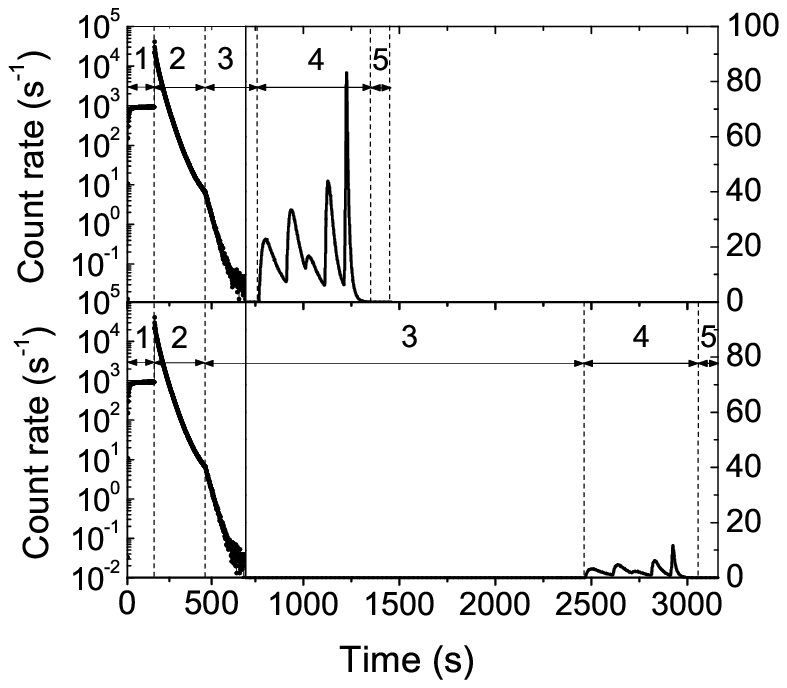}
\caption{\label{fig2} Time diagrams of the storage cycle for two
different holding times in a narrow trap. 1--filling, 160 s
(time of trap rotation (35 s) to monitoring position is included);
2--monitoring, 300 s; 3--holding, 300 s or 2000 s (time of trap
rotation (7 s) to holding position is included); 4--emptying, with five
periods of 150, 100, 100, 100, and 150 s (with time of trap rotation
2.3, 2.3, 2.3, 3.5, and 24.5 s to each position 
included); 5--measurement of background, 100 s.}
\end{figure}

\begin{figure}
\includegraphics[width=3.5in]{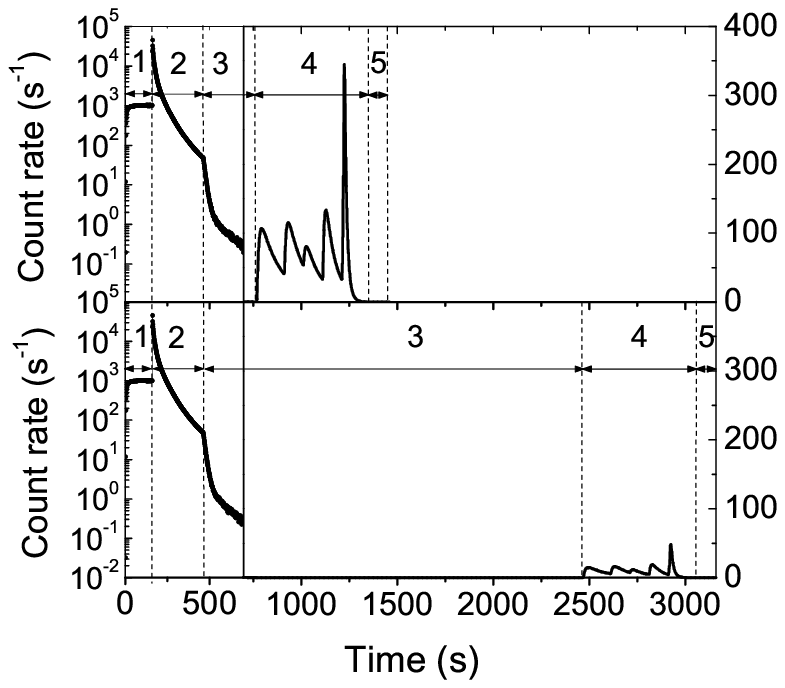}
\caption{\label{fig3} Time diagrams of the storage cycle for two
different holding times in a quasispherical trap. 1--filling, 160 s
(time of trap rotation (35 s) to monitoring position is included);
2--monitoring, 300 s; 3--holding, 300 s or 2000 s (time of trap
rotation (7 s) to holding position is included); 4--emptying, with five
periods of 150, 100, 100, 100, and 150 s (with time of trap rotation
2.3, 2.3, 2.3, 3.5, and 24.5 s to each position 
included); 5--measurement of background, 100 s.}
\end{figure}

A typical measurement of the counting rate at the detector in a data 
cycle with a narrow trap is shown in Fig.~\ref{fig2}.
Figure~\ref{fig3} shows a similar measurement for data cycle taken 
with a quasispherical trap.

At the beginning of a measurement cycle, the trap is in the
``window-down'' position, as it is being filled with UCN. After this,
the trap is rotated into the monitoring position $\theta=30^\circ$,
in which the trap window height is approximately 10 cm lower than
in the ``window-up'' position. The filling process can
be monitored by the UCN detector (see Fig.~\ref{fig1}) through
a slit in the selector valve. After the trap has been rotated into
the monitoring position, the selector valve is switched into the
position in which UCN are registered. The trap is maintained in
the monitoring position for 300 s. The neutrons that are in volume
between the trap walls and the cryostat walls are registered by the detector.
During this period, neutrons
with energies higher than the gravitational barrier leave the trap
(see Fig.~\ref{fig2}). The trap is rotated into the ``window-up''
position after the monitoring period. The procedure of the neutron filling
and preparation of the UCN spectrum takes about 700 s; in the left axis
of Fig.~\ref{fig2} the counting rate is presented on a logarithmic scale:
the counting rate for the subsequent procedures (700--3000 s)
are presented with a greater detail on a linear scale (the right axis of
Fig.~\ref{fig2}).
After a short (upper half of Fig.~\ref{fig2}) or long (lower part of
 Fig.~\ref{fig2}) storage time, the trap is rotated in steps 
through five positions, and in each of these positions it is kept
fixed for 100--150 s, so that the UCN can be registered.
The neutrons detected after the each rotation have different mean energies.
After all the UCN have left the trap, the background count rate is measured.

The angular positions of the quasispherical trap window
 and the mean values of the UCN
energies in each discharge are given by\\
$\theta=30^\circ$ (monitoring position);\\
$\theta=40^\circ,  E_{UCN}=62.3$ neV;\\
$\theta=50^\circ,  E_{UCN}= 56.4$ neV;\\
$\theta=55^\circ,  E_{UCN}= 52.3$ neV;\\
$\theta=70^\circ,  E_{UCN}= 43.2$ neV;\\
$\theta=180^\circ, E_{UCN} = 26.7$ neV.

The angles are given with respect to the vertical up direction
and were chosen in such a way that there will be an equal fraction
of the spectrum emptied at each position (Unfortunately, the
third portion was not successfully optimized, which is clearly
obvious from Fig.~\ref{fig2}.) 
The diameter of the narrow trap is smaller compared to the
quasispherical trap. Therefore, slightly different angles for
a narrow trap were chosen to obtain the same energy intervals.

\section{Methods of extrapolation to the neutron lifetime}
\subsection{Basic relations}
The storage time for UCN in a system with a loss vessel is given by:
\begin{equation}
\label{eq5}
\tau _{st}^{-1} =\tau _n^{-1} +\tau _{loss}^{-1}
\end{equation}
Here, the total UCN loss $\tau _{st}^{-1}$ comprises
two terms, namely, the probability of neutron $\beta $ decay $\tau
_n^{-1} $, and the probability $\tau _{loss}^{-1} $ of UCN losses.
The success of the experiment is related to the fact that $\tau
_{loss}^{-1} $ is small compared to $\tau _n^{-1} $. In our
experiment, $\tau _{loss}^{-1} $ is 1{\%} from $\tau _n^{-1}$.
Therefore the measurement of $\tau _{loss}^{-1}$ with 10{\%}
accuracy gives the opportunity of  $\tau _n^{-1}$ determination
with the 0.1{\%} or $\pm 0.8$ s accuracy.

Although the task of calculating $\tau _{loss}^{-1}$ with 10{\%}
accuracy looks rather simple, it should be done very carefully.
Therefore, we will consider all stages of the $\tau _{loss}^{-1}$
calculation  in succession and in detail.

Let us imagine monoenergetic UCN captured in a trap with volume
$V$ and surface area $S$. For simplicity, we will neglect gravity
for the moment. The total number of UCN in the trap will be
\begin{equation}
\label{eq6}
N(E)= V \cdot \rho (E),
\end{equation}
where $\rho (E)$ is the UCN number density, which depends on the
UCN energy $E$. The number of UCN lost in the trap per second will
be
\begin{equation}
\label{eq7}
\frac{dN(E)}{dt}=\left( {1 \mathord{\left/ {\vphantom {1 4}} \right.
\kern-\nulldelimiterspace} 4} \right)v(E)\rho (E)S \mu(E),
\end{equation}
where $v(E)$ is the UCN velocity and the function $\mu
(E)$ gives the UCN losses from reflection, which
depends on the UCN energy and on some properties of the trap
walls. (Note that $\frac{1}{4}v\rho S $ is the flux of UCN on
the trap surface.)
Using Eqs. (\ref{eq6}) and (\ref{eq7}) one can
write the expression for loss probability on the trap walls as
\begin{equation}
\label{eq8}
\tau _{loss}^{-1} (E)=\frac{dN(E)}{N(E)dt}=\frac{\left( {1 \mathord{\left/
{\vphantom {1 4}} \right. \kern-\nulldelimiterspace} 4}
\right)v(E)\rho \mbox{(}E\mbox{)}S\mu (E)}{V \cdot \rho
\mbox{(}E\mbox{)}}
\end{equation}
or
\begin{equation}
\label{eq9}
\tau _{loss}^{-1} (E)=\frac{S}{4V}\cdot v(E)\cdot \mu (E)
\end{equation}
If we define $l$ as a mean free path of UCN in the trap 
($l ={4\cdot V} \mathord{\left/ {\vphantom {{4\cdot V} S}} \right.
\kern-\nulldelimiterspace} S$) and $f(E)$ as the UCN collision
frequency ($f(E)=v(E)/l$), Eq. (\ref{eq9}) for the loss probability
can be simplified to
\[
\tau _{loss}^{-1} =f\left( E \right)\cdot \mu \left( E \right).
\]
By assuming that the UCN are reflected from a potential step with
real (U$_{0})$ and imaginary (W) parts, the UCN losses from  reflection
can be represented in the following well-known form \cite{Ign90}:
\begin{equation}
\label{eq10}
\mu (y)=\frac{2\eta }{y^2}\cdot (\arcsin y-y\sqrt {1-y^2} )\approx \left\{
{{\begin{array}{*{20}c}
 {\pi \eta , y\to 1} \hfill \\
 {\frac{4}{3}\eta y, y<<1} \hfill \\
\end{array} }} \right.
\end{equation}
where $\eta =W/U_0 =b'/b$ is the loss factor determined by the ratio of
the imaginary to real parts of the potential or the scattering
amplitudes, and $y=(E/U_0 )^{1\mathord{\left/ {\vphantom {1 2}} \right.
\kern-\nulldelimiterspace} 2}$.

In Eq.(\ref{eq10}), the energy UCN loss function has been
averaged over the isotropic distribution of the incidence angles.
Using the optical theorem, we can write the imaginary part of
the scattering amplitude in the following form \cite{Ign90}:
\[
b' = \frac{{\sigma _{abs}  + \sigma _{upscat} (T)}}{{2\lambda }}
\]

The capture and inelastic-scattering cross sections are
proportional to the neutron wavelength $\lambda$, with the result
that neither $b'$ nor $\eta$ depends $\lambda$ or the neutron
energy $E$. However, the loss factor is temperature dependent,
$\eta=\eta(T)$, owing to the temperature dependence of the
inelastic-scattering cross section $\sigma _{upscat} (T)$.

It looks very useful to rewrite the right-hand side of Eq.(\ref{eq9})
as the product of two factors, one depending only on
the trap temperature and the other only on the UCN energy:
\begin{equation}
\label{eq11}
\tau _{loss}^{-1} (T,E)=\eta (T)\cdot \gamma( E )
\end{equation}
where $\gamma(E)$ is the normalized loss rate which depends on the
UCN energy and the trap dimensions.

Taking into account Eq.(\ref{eq11}) we can rewrite Eq.(\ref{eq5}) as
\begin{equation}
\label{eq12}
\tau _{st}^{-1} (E)=\tau _n^{-1} +\eta(T)\cdot \gamma(E)
\end{equation}
Now we are ready to explain the principle of how to extract the
neutron lifetime value from the experimental storage times. For this
purpose, we have to measure the UCN storage time in two experiments
($\tau_1$ and $\tau_2$). The normalized loss rates $\gamma _1 $
and $\gamma _2 $ should be different in these measurements. Using
Eq.(\ref{eq12}) for the total UCN loss probability, namely,
\begin{equation}
\label{eq13}
\tau _1^{-1} =\tau _n^{-1} +\eta \gamma_1(E)
\end{equation}
\begin{equation}
\label{eq14}
\tau _2^{-1} =\tau _n^{-1} +\eta \gamma_2(E)
\end{equation}
we find that
\begin{equation}
\label{eq15} \tau _n^{ - 1} = \tau _1^{ - 1}  - \frac{{\tau _2^{ -
1}  - \tau _1^{ - 1} }} {{{{\gamma _2 (E)} \mathord{\left/
 {\vphantom {{\gamma _2} {\gamma _1   - 1}}} \right.
 \kern-\nulldelimiterspace} {\gamma _1 (E)  - 1}}}}
\end{equation}

One can see that Eq.(\ref{eq15}) contains only the ratio of
the normalized loss rates $\gamma_1$ and $\gamma_2$. Different values
of the UCN normalized loss rate $\gamma$ can be obtained by using
traps of different dimensions and/or different values of the UCN
energy. Therefore, we can use either energy extrapolation or size
extrapolation.

For the energy extrapolation we will use one trap and will
capture the UCN of different energy. Using Eqs.(\ref{eq13}) and 
(\ref{eq14}) we can extrapolate to zero losses to obtain
 $\tau _n^{-1}$. But the rest of the dependence of
 $\tau _n^{-1}$ on the function $\mu(E)$
exhibits some problem. Actually the real function $\mu (E)$
may differ somewhat from that of the adopted form (\ref{eq10}),
which was calculated for the ideal potential step.

To exclude the effect of the energy dependence $\mu (E)$ on the
result, we can extrapolate to zero losses by using data on
the storage of UCN with the same mean energy in traps of
different dimensions. This method is called the size
extrapolation method. In this case the ratio of the normalized loss
rate $\gamma_2$ to $\gamma_1$ depends only on the trap sizes:
\begin{equation}
\label{eq17} {\gamma _2 } \mathord{\left/ {\vphantom {{\gamma _2 }
{\gamma _1 }}} \right. \kern-\nulldelimiterspace} {\gamma _1
}={\frac{S_2 }{V_2 }} \mathord{\left/ {\vphantom {{\frac{S_2
}{V_2 }} {\frac{S_1 }{V_1 }}}} \right.
\kern-\nulldelimiterspace} {\frac{S_1 }{V_1 }}
\end{equation}

Thus, the effect of any energy dependence is excluded entirely.
Clearly, this is valid only in the absence of gravity.

In a gravitational field UCN kinetic energy ${E}'$ becomes
dependent on the height: ${E}'=E-mgh$, where $h$ is the height
measured from the bottom of the trap and $E$ is the UCN energy at
$h=0$. So the gravity-free equation (\ref{eq8}) has to be
modified. We have to replace $E$ by $E-mgh$ and to integrate
with respect to $h$ on the trap surface in the numerator and on the
trap volume in the denominator:
\begin{eqnarray}
\label{eq18}
\tau _{loss}^{-1} (E)=\eta \cdot \gamma \left( E \right)=\nonumber\\
\frac{\int\limits_0^{E \mathord{\left/ {\vphantom {E {mg}}} \right.
\kern-\nulldelimiterspace} {mg}} {\mu (E-{h}')\cdot v(E-{h}')\cdot
\rho (E-{h}')dS(h)} }{4\int\limits_0^{E \mathord{\left/ {\vphantom {E {mg}}}
\right. \kern-\nulldelimiterspace} {mg}} {\rho (E-{h}')dV(h)} },
\end{eqnarray}
where ${h}'=mgh$. The UCN number density is proportional to 
$\sqrt {{\left({E-mgh} \right)} \mathord{\left/ {\vphantom {{\left(
 {E-mgh} \right)} E}}\right. \kern-\nulldelimiterspace} E} $,

In reality, when gravity is present, complete exclusion of $\mu
(E)$ is impossible because of the integral nature of Eq.(\ref{eq18}).
However, the residual effect of the dependence $\mu (E)$ on the
neutron lifetime is negligible. 
It will be shown further that in our experiment on
the neutron lifetime measurement, the contribution from
uncertainty of this dependence does not exceed 0.144~s,
whereas the statistical accuracy of the
measurement was 0.7~s. To demonstrate this fact the UCN normalized loss
rates $\gamma$ were calculated for a series of model dependencies
$\mu (v)$:

1. $\mu (v)\sim$  constant,

2. $\mu (v)\sim v / v_{lim}$,

3. $\mu (v)\sim \left( v / v_{lim} \right)^2$,

4. $\mu (v)\sim \left( v / v_{lim} \right)^3$,\\
where $v_{lim}$ is the trap boundary velocity.
A coefficient in these dependencies is not important
because it is outside the integral in Eq.(\ref{eq18}) and is canceled in
Eq.(\ref{eq15}). Then, the values of
the neutron lifetime were obtained by the size extrapolation
method for each corresponding model dependency $\mu _i (v)$.
The differences between the neutron lifetime values for
$\mu(v)$ from Eq.(\ref{eq10}) and for the model dependencies
$\mu_{i}(v)$ are as follows:

1. $-0.158$ s,

2. $-0.022$ s,

3. $+0.1$ s, and

4. $+0.217$ s.\\
The linear dependence $v / v_{lim}$ is the most
similar to the theoretical dependence $\mu(v)$, especially for small
velocities.
The mean square value of the numbers listed is $0.144$~s and can be
used as an estimation of the uncertainty of the neutron lifetime value owing to
the uncertainty of the $\mu(v)$ function shape (see Table~\ref{table}). In
spite of the fact that the model dependencies $\mu_{i}(v)$ are
very different, their influence on the result of the extrapolation
to the neutron lifetime is rather weak. Therefore, the size
extrapolation method based on the idea of using two traps makes it
possible to reduce substantially systematic errors that are
caused by the uncertainty of our knowledge about the function
$\mu(E)$.

The effect from uncertainty of the dependence $\mu (E)$ on the neutron
lifetime was also studied for the case of the energy extrapolation method.
The differences between the neutron lifetime values for $\mu(v)$ from 
Eq.(\ref{eq10}) and for the model dependencies $\mu_{i}(v)$ are as follows:

1. $-0.83$ s,

2. $0.76$ s,

3. $-1.93$ s, and

4. $-2.73$ s.\\
As expected, the dependence of the result of the energy
extrapolation method on the $\mu(v)$ function shape is rather
strong, going mainly to systematically low value of neutron
lifetime. Therefore, we used the result of the size extrapolation as
the final value of the neutron lifetime.

\begin{figure}
\includegraphics[width=3.5in]{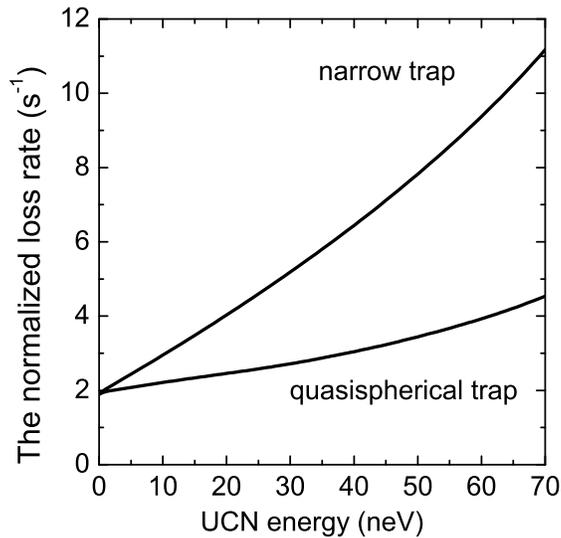}
\caption{\label{fig4} The UCN normalized loss rate as a function of
UCN energy for narrow cylindrical and wide quasispherical traps.}
\end{figure}

If we have more than two measurements, the neutron lifetime can be
obtained by a linear regression of Eq.(\ref{eq12}) with
$\tau_{n}^{-1}$ and $\eta$ as free parameters. The UCN loss factor
$\eta $ proves to be equal to the tangent of the slope angle of
the extrapolation line. The UCN normalized loss rate $\gamma(E)$
with the trap walls can be computed according to Eq.
(\ref{eq18}). The results of calculation of the normalized loss
rate are presented in Fig.~\ref{fig4}.

The UCN storage time $\tau _{st}^{-1}$ can be calculated if we
know the measured number of neutrons that remain in the trap after
two different storage times:
\begin{equation}
\label{eq19}
\tau _{st}  = \frac{{t_2  - t_1 }}{{\ln ({{N_1 } \mathord{\left/
 {\vphantom {{N_1 } {N_2 )}}} \right.
 \kern-\nulldelimiterspace} {N_2 )}}}},
\end{equation}
where $N_1$ and $N_2$ are the numbers of neutrons left in the trap
after storage times $t_1$ and $t_2$, respectively. There is no
need to know the efficiency of the UCN detector, the probability
of UCN losses as the neutrons travel along the neutron guide, and
so forth, since Eq.(\ref{eq19}) uses only the ratio of the
numbers of neutrons.

\subsection{\label{ñalculation loss rate}Calculation of normalized 
 loss rate for the broad UCN spectrum in the trap}
 
Previously it was demonstrated how one can calculate the
normalized loss rate $\gamma(E)$ and fulfill the size extrapolation
to the neutron lifetime for a the monoenergetic UCN spectrum. But
in practice we have dealt with a rather broad spectrum.

To calculate the normalized loss rate for a broad UCN
spectrum one has to integrate the Eq.(\ref{eq18}) over the
real spectral distribution of UCN in the trap. Hence the task of
calculating the normalized loss rate for a broad spectrum leads
to the task of spectrum measurement.

There are two ways to determine the UCN spectrum in the trap:
1. by direct measurement of the differential UCN spectrum by means of the 
step-by-step rotation of the UCN trap as was previously described and 
2. by measuring the integral UCN spectrum and then calculating its
differential form. In the first case, each portion of the spectrum will
be measured in a different time. Therefore, it is necessary to introduce
a correction for the UCN decay and losses. In the second case, introduction
of this correction is not required. We used the second way, measuring the
number of neutrons in the trap by means of the trap rotation
to the down position from the different monitoring positions.

\begin{figure}
\includegraphics[width=3.5in]{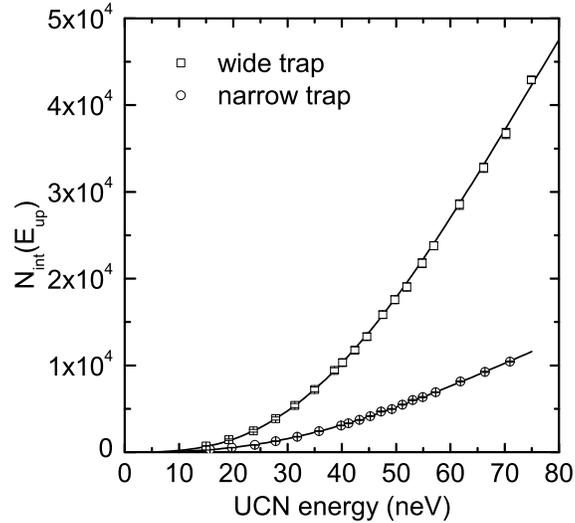}
\caption{\label{fig5} The integral UCN spectra. The open squares
represent the results of measurements for a quasispherical trap,
and the open circles are the results of measurements for a cylindrical
trap. The results of the least squares fitting are shown by solid
line.}
\end{figure}

The differential spectrum of UCN captured into the trap, $N(E)$,
can be calculated by the so-called integral spectrum $N_{int}
(E_{up} )$. The integral spectra for each trap were
measured during a special experiment and the results are presented in
Fig.~\ref{fig5}.

The differential spectrum was taken as
\begin{equation}
\label{eq20}
N(E)=\rho _0 \cdot \rho (E)\cdot V(E)\cdot \exp (-\alpha E),
\end{equation}
where $\rho _0 $ is the UCN density in phase space, $V(E)$ is the
effective volume for UCN with energy $E$ in gravitational field,
$E$ is the UCN energy at the trap bottom, $\alpha$ is the deformation
parameter of the Maxwell spectrum, and $\rho (E)$ is Maxwell spectrum of
UCN ($v^2dv \sim \sqrt{E}dE)$.

\begin{figure}
\includegraphics[width=3.5in]{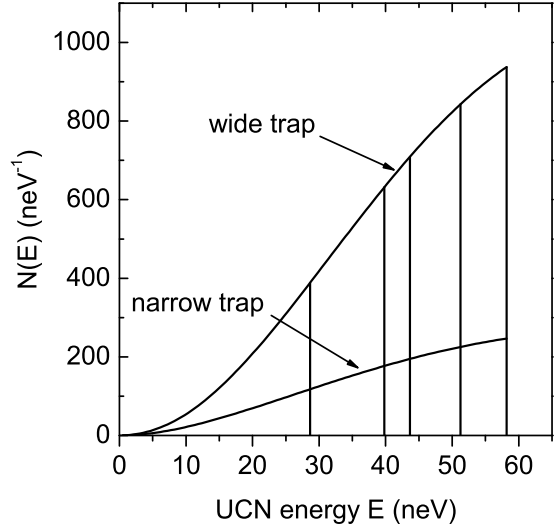}
\caption{\label{fig6} The differential UCN spectra in 
quasispherical and cylindrical traps. The position of each
vertical line is defined by the height of the gravitational
barrier for corresponding emptying.}
\end{figure}

The function $V(E)$ depends on the trap configuration and was
calculated from
\begin{equation}
\label{eq21}
V(E)=\int\limits_0^{E \mathord{\left/ {\vphantom {E {mg}}} \right.
\kern-\nulldelimiterspace} {mg}} {\sqrt {\frac{E-mgh}{E}} } \cdot A(h)dh,
\end{equation}
where $A(h)$ is the cross-sectional area of the trap at height
$h$. The experimental values of integral spectra for each 
trap were fitted by the least squares method with a function that
equals the integral from $N(E)$ [Eq.\ref{eq20}]:
\begin{equation}
\label{eq22} N_{int} (E_{up} )=\rho _0 \int\limits_0^{E_{up} }
{V(E)\cdot \sqrt E \cdot} \exp (-\alpha E)dE.
\end{equation}
The values of parameters $\alpha $ and $\rho _0 $ were determined
as a result of the least squares fitting, which is shown in
Fig.~\ref{fig5} by a solid line. The differential spectra of UCN
captured in the trap were calculated according to Eq.(\ref{eq20}) and
are presented in Fig.~\ref{fig6} for quasispherical and
cylindrical traps. The shape of the spectra depends mainly on the
parameter $\alpha$. One can determine the differential spectrum by
a simple differentiation of the measured integral spectrum. This method
looks more straightforward, as it does not need any model function
such as Eq.(\ref{eq20}), but direct differentiation introduces some
additional uncertainties. Finally, both methods result in
approximately the same UCN spectrum uncertainty in
Table~\ref{table}.

The main procedure of the storage time measurement uses five
successive emptying stages, as shown in Figs.~\ref{fig2} and 
\ref{fig3}. Let us
discuss now the method of the spectrum calculation for each
successive emptying. Each emptying differs by the height of the
gravitational barrier for UCN, which is defined by the height of
the low edge of the trap window relative to the trap bottom. The
neutrons with energies higher than this gravitational barrier can
leave the trap. Suppose for the moment that all such neutrons
leave the trap. This is strictly valid if emptying times are
infinitely long. In this case, the spectrum of captured UCN
should be divided into intervals of UCN energy, as shown in
Fig.~\ref{fig6} by vertical lines. The position of each line is
defined by the height of the gravitational barrier for
the corresponding emptying.

The values of the normalized loss rate $\gamma _i $ for energy
interval $i$ can be calculated as follows:
\begin{equation}
\label{eq23}
\bar {\gamma }_i =\frac{\int_{E_{i-1} }^{E_i } {\gamma (E)\cdot N(E)dE}
}{\int_{E_{i-1} }^{E_i } {N(E)dE} },
\end{equation}
where $E_i$ is the bound of energy interval $i$, with $E_0 =0$.

\begin{figure}
\includegraphics[width=3in]{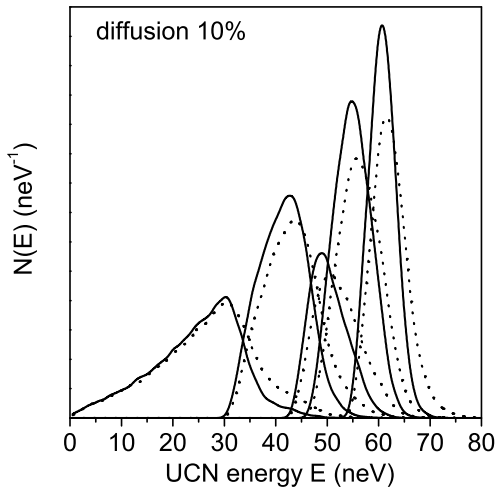}
\caption{\label{fig7} Monte Carlo simulation of UCN differential
spectra in a cylindrical trap (solid line) and in a quasispherical
trap (dotted line) for each of five emptyings. Spectra are 
normalized to unity. The probability of UCN
diffuse scattering used in the simulation is 10{\%}.}
\end{figure}

However, in reality the emptying times are finite. Hence some neutrons
with energies greater than the gravitational barrier
do not leave the trap. The spectrum calculation should take
into account five successive emptyings with finite time and the
rotation of the trap with finite speed. The best method for such
a calculation is a Monte Carlo simulation. The results of such a Monte
Carlo simulation for narrow cylindrical and quasispherical traps
are presented in Fig.~\ref{fig7}. Compared to the spectra for the interval
method (see Fig.~\ref{fig6}) one can observe the shape change
and shift of mean energies to the high end.

The normalized loss rate $\gamma _i $ for the $i$-emptying can be
calculated according to
\begin{equation}
\label{eq24} \bar {\gamma }_i =\frac{\int_0^\infty {\gamma
(E)\cdot N_i (E)dE} }{\int_0^\infty {N_i (E)dE} },
\end{equation}
where $N_i (E), i=1\ldots 5$, is the UCN spectrum of the $i$-emptying as
shown in Fig.~\ref{fig7}.

A comparative analysis of the extrapolation of the experimental data to the
neutron lifetime using Eqs.(\ref{eq23}) and (\ref{eq24})
shows that the difference in the extrapolated value $\tau _n $ is only
$0.15$ s.

This method of calculating $\gamma _i $ on energy intervals was
used during the experimental data treatment, but the energies
$E_{i}$ and $E_{i-1}$ in Eq.(\ref{eq23}) were corrected for
incomplete emptying. So, the effect of the mean energy shift was
taken into account. Note that this shift of the mean energy is
slightly different for cylindrical and quasispherical traps 
because of the different ratio of the window area to the
trap volume for these traps. One can observe this effect in 
Fig.~\ref{fig7}. However, the influence of this effect is
rather small. To check it, we used the same spectrum for the calculation
of $\gamma$ values in cylindrical and quasispherical traps.
The extrapolated value of the neutron lifetime was shifted by 0.23 s
compared to the extrapolation where we used the native spectra for
different traps.

\section{Neutron lifetime measurements}
\subsection{Low-temperature perfluoropolyether}

For this experiment, a new type of material -- a low-temperature, fully
fluorinated polymer -- was proposed \cite{Pok99} and used for
coating trap walls. The chemical formula of the substance --
perfluoropolyether (PFPE)\footnote{The product was manufactured 
by the Perm branch of the Russian Scientific Center for Applied
Chemistry.} -- is $CF_3 O(CF_2 O)_n (CF_2 CF_2 O)_m
(OCF_2 CF_2 O)_l CF_3$ with $n \approx 30.3$, $m \approx 1.5$, $l
\approx 0.2$, and molecular weight M=2350; its vapor pressure at room
temperature was measured to be about $1.5 \times 10^{-3}$ mbar and the
pour temperature was measured to be about $-100^\circ$C \cite{Pok03}. This
substance was deposited on the trap surface by evaporation in vacuum.

The PFPE contains only C, O, and F, so its neutron-capture
cross section is small. As a result of a preliminary study
of several types of PFPE, it was found \cite{Ste02} that the
quasielastic and inelastic UCN scattering in PFPE for  
$T < -120^\circ$C is much weaker than in ordinary Fomblin at room
temperature. Quasielastic UCN scattering is completely suppressed
for $T < -120^\circ$C \cite{Ste02}, and because of the neutron inelastic
scattering the expected UCN loss factor $\eta $ amounts to roughly
$2 \times 10^{-6}$ \cite{Pok03}.

Before the beginning of the PFPE deposition procedure, a spherical
vaporizer with tiny holes was heated to 140$^\circ$C with an electric
heater. Then gaseous helium was utilized to force three cubic
centimeters of liquid into the vaporizer's chamber along a
vertical tube. The deposition of the PFPE was done by evaporation with the
evaporated substance from the vaporizer being settled and frozen onto the
inner walls of the trap cooled to $-150^\circ$C. To achieve
homogeneity, the vaporizer was moved up and down.

\subsection{Study of the coating properties of PFPE}

To check the quality of the PFPE film, a copper trap with
a titanium coating was employed. Titanium has a negative
scattering length and does not generate a reflecting potential for
UCN. Ultracold neutrons cannot be stored in such a trap if the
titanium coating is not covered with a layer of PFPE. The trap was a
50-cm-long cylinder with a diameter of 76 cm. A stable storage
time $\tau_{st}=869.0 \pm $0.5~s was achieved after several
depositions of PFPE (with a total thickness of 15 $\mu $m)
with the temperature of the trap walls varying from $-140^\circ$ to
$-150^\circ$C and after a single heating-cooling cycle in which
the trap temperature was first raised to room temperature and then
brought down to T = $-160^\circ$C. The repetition of the thermal cycling
had no effect on the UCN storage time in the trap. It is quite
possible that at room temperature the PFPE filled all the gaps and
cracks in the trap walls and formed a perfect surface. In
addition, PFPE was degassed in the thin layer at room temperature.
Such a coating is extremely stable, and no essential variation in
the storage time was detected during the eight-day period of
observation. Subsequent depositions had no effect on the value of
the UCN storage time in the trap.

\begin{figure}
\includegraphics[width=3.5in]{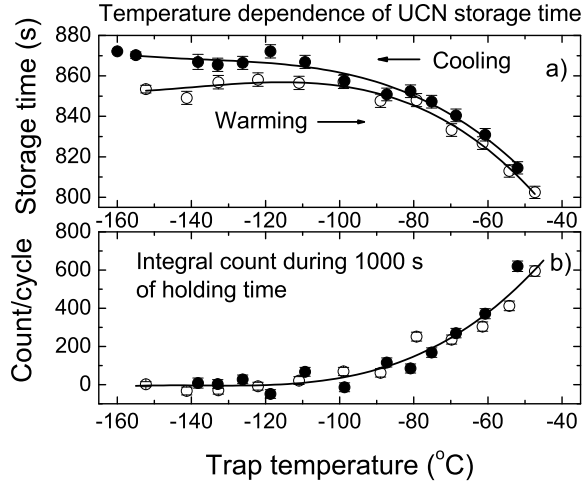}
\caption{\label{fig8} Temperature dependence of (a) the UCN
storage time in the course of warming (empty circles) and the
cooling again (full circles) after the first evaporation of PFPE
at T = --155$^\circ$C and (b) the integral count for a 1000-s
storage period after subtraction of background and count rate
of UCN that have energy higher than the gravitational barrier 
(see Fig.~\ref{fig10} and Sec.~\ref{Monte Carlo simulation})}.
\end{figure}

The traps used in the final measurements  (a quasispherical trap
and a narrow cylindrical one) were coated with beryllium.
Since beryllium constitutes a good reflector of ultracold
neutrons, the trap walls can be cooled to even lower temperatures,
since the appearance of any microcracks in the coating has a
small influence on the UCN lifetime in the trap. Using a
beryllium-coated trap, we studied the temperature dependence of
the storage time for a quasispherical trap coated with PFPE
[Fig.~\ref{fig8}(a)].

PFPE was frozen on the trap wall at T = $-155^\circ$C, then the trap
was slowly heated to T = $-50^\circ$C, and finally cooled down
again to T = $-160^\circ$C. In this way, we covered up the layer
defects with oil when it was fairly liquid. After this temperature
cycling was completed, the measured storage time turned out to be
longer, 872.2$\pm $0.3~s, than immediately after sputtering,
850$\pm $1.8~s. Repeated warming and cooling cycles did not change
the UCN storage time of the traps (see Fig.~\ref{fig11}).

This procedure (heating after evaporation at T = $-160^\circ$C) is
extremely important because the oil in a liquid form coats all
surface defects because of the effect of the surface tension of PFPE.

By comparing the storage time of the trap with a titanium substrate and
with a beryllium substrate we can estimate the part uncoated by
the PFPE surface. Bearing in mind that the titanium and beryllium
traps differ in dimensions, we calculated the difference between
the expected and measured storage times for the titanium trap,
which was 1.9$\pm $0.6~s. This is equivalent to an uncovered part
of the surface area of the titanium trap equal only to
($4.4\pm 1.3)\times 10^{-7}$. Hence, the reproducible PFPE coating
can be obtained irrespective of the material and shape of the
trap. For this reason, there is no need to examine the different
loss factors $\eta $ for various traps with a beryllium sublayer
under a PFPE coating.

\subsection{Study of quasielastic UCN scattering}
In the course of investigations that used a beryllium-coated trap,
the quasielastic UCN scattering by PFPE was studied.

In our installation, the principle of the gravitational valve was used.
A gravitational valve cannot store UCN with energy more than the
gravitational barrier. If UCN during the storage take up additional
energy they will leave the trap and reach the detector.

Figure~\ref{fig8}(b) shows the number of ultracold neutrons that
leave the trap during a 1000-s (761-1760 s) storage period as
a function of the trap temperature.
Here the background and count rate of UCN that
have energy higher than the gravitational barrier  are already
subtracted (see Fig.~\ref{fig10} and Sec.~\ref{Monte Carlo simulation}).
In the process,  we observed an additional counting rate (on top of
the background noise),
which fell off exponentially with the passage of the time of UCN
storage in the trap. The additional counting rate appears because
the UCN acquire energy in their quasielastic scattering by PFPE.
These neutrons leave the trap, which drives the counting rate at
the detector upward. The counting of quasielastically scattered UCN
becomes indistinguishable against the background (i.e., it disappears)
as T $< -120^\circ$C. This result is in qualitative
agreement with that of measuring the quasielastic UCN scattering
by PFPE studied in our previous work \cite{Ste02}.

\begin{figure}
\includegraphics[width=3.5in]{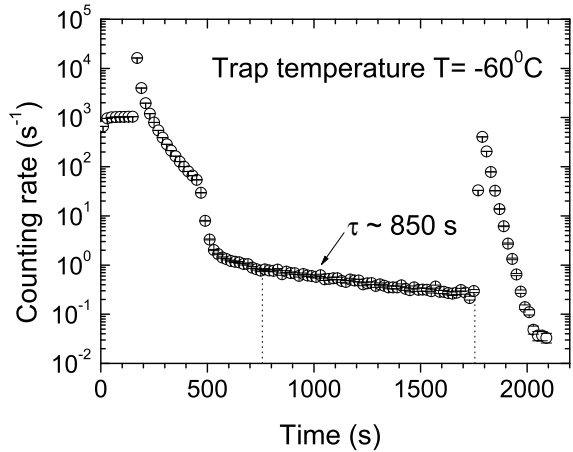}
\caption{\label{fig9} Time diagram of the neutron storage cycle in
a quasispherical trap at T = $-60^\circ$C.
The exponential time of emptying of quasielastically
scattered UCN is defined by the UCN storage time in the trap.
}
\end{figure}

\begin{figure}
\includegraphics[width=3.5in]{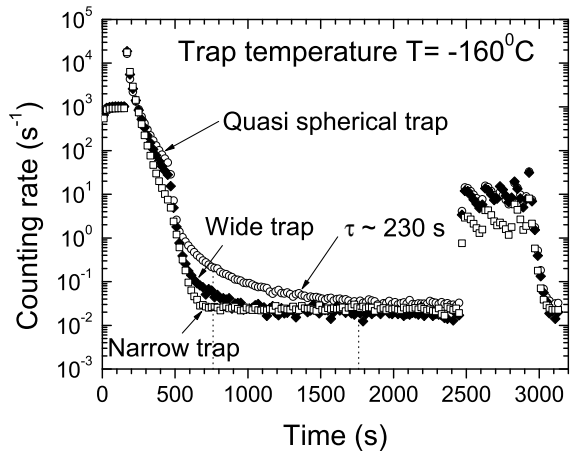}
\caption{\label{fig10} Time diagram of the neutron storage cycle
in narrow cylindrical, wide cylindrical, and quasispherical trap
at T = $-160^\circ$C.
The exponential time of emptying of UCN that have
energy higher than the gravitational barrier depends on the shape
of the trap (see Sec.~\ref{Monte Carlo simulation}). For the
quasispherical trap the exponential time of emptying of UCN that have
energy higher than the gravitational barrier is $\sim230$~s. The time diagram
during a storage period has no dependence on exponential times
such as the storage time in the trap ($\sim800-850$~s), because
the process of quasielastic UCN scattering is suppressed at
T = $-160^\circ$C.
}
\end{figure}

Here, this process was studied in more detail, particularly
at temperature T = $-160^\circ$C. At  T =$-60^\circ$C we find 
reasonable agreement with the result of our previous work
\cite{Ste02}. The illustration of the process of UCN upscattering at 
T = $-60^\circ$C is shown in Fig.~\ref{fig9}. We can see
the clear definite exponent with the UCN storage time in the trap.

The most important studies have been carried out at a trap
temperature T = $-160^\circ$C. Figure~\ref{fig10} shows the counting
rate of the detector during UCN storage in the traps with 
T = $-160^\circ$C. Just after the trap rotation in the storage
position (hole up) we can see the counting rate of UCN whose
energy is above the gravitational barrier. The exponential time is
different for different traps (narrow, wide cylindrical, and
quasispherical) because of the different ratio of the window area to the
trap volume. The exponential time of counting rate is in
a reasonable agreement with Monte Carlo calculations. For the narrow
trap this process is the fastest (about 30 s); therefore this
curve is the most preferable for the analysis of UCN upscattering.

The counting rate of quasielastically scattered UCN should be
proportional to the number of UCN in the trap, UCN collision
frequency with the trap walls, and probability of UCN upscattering.
The analysis shows that the counting rate during a storage period
contains no counts with an exponential time like the storage time 
(about 800--850~s). But this conclusion is valid with 
finite uncertainty. Therefore, we can only calculate the upper limit
for the probability of UCN upscattering with energy transfer about 20
neV. This upper limit is $6\times 10^{-9}$ per collision. From
this upper limit of UCN probability of leakage at storage in the
trap we can also infer the upper limit of the correction to the neutron
lifetime, which is less than 0.03~s. It should be mentioned that
the estimation is valid for any process of low-energy transfer owing
to the interaction with the trap surface as well as the interaction inside the
trap volume.

Thus, it turned out that for T~$<-120^\circ$C, the quasielastic UCN
scattering can be ignored. Our measurements were made at T =
$-160^\circ$C to guarantee that quasielastic scattering does not
affect the results.

The effect of emptying UCN that have energy higher than the
gravitational barrier for the quasispherical trap is insignificant
too. That part of UCN that have energy higher than the gravitational
barrier is $10^{-3}$ (i.e., on the level of statistical error).
Moreover, our data were corrected for this effect. 

\subsection{Study of the stability and reproducibility of coating}

\begin{figure}
\includegraphics[width=3.5in]{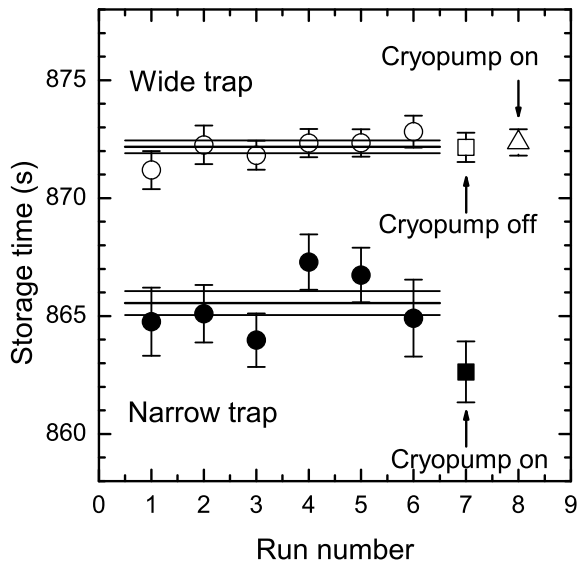}
\caption{\label{fig11} Demonstration of the stability of a PFPE
coating during measurements. The UCN storage times for the wide
and narrow traps differ because of the different frequencies of
UCN collisions with the trap walls.}
\end{figure}

The stability and integrity of the coatings on various traps
constitute the most important conditions for the use of the size
extrapolation method in measuring the neutron lifetime. Therefore,
the quality of the PFPE coating was checked many times during
measurements. Figure~\ref{fig11} gives eight results of
the measurements of the neutron storage time for a quasispherical
trap and seven analogous results for a narrow trap. The
measurements were carried out after new depositions, heating, and
cooling with a subsequent new deposition, etc. After a
liquid-helium cryogenic pump was mounted near the storage volume,
the trap's vacuum was improved from $5 \times 10^{-6}$ to $3\times
10^{-7}$ mbar. The storage time during the experiment on the
measurement of the neutron lifetime agrees, within approximately one
second, for the wide trap and, within marginally broader limits,
for the narrow trap. This proves that the PFPE coatings are stable
and reproducible for various traps.

Thus, assuming that the substances
used for coatings in various UCN traps were identical and taking
into account the exceptionally high coating properties of a
Fomblin oil (owing to its surface tension) and the
absence of coating degradation, we believe that the same loss
factor $\eta $ can be used for different traps.

The level of statistical accuracy of our experiment agrees with
the requirement that the loss factor $\eta $ within about 5{\%} is
the same for different traps. However, test experiments with the
titanium trap have convinced us that this requirement is met with
a much higher accuracy, because any defects of coating of the
titanium trap would cause the UCN storage times to be very unstable.

The reasons for using the same loss factor $\eta $ for different traps
are the following:
\begin{itemize}
	\item  The coating of the traps was made of the same
	 substance; moreover, the substance was taken from the same bottle.
	\item There is a high coating capacity of the oil in liquid form owing
	 to the effect of the surface tension of PFPE.
	\item There is an absence of degradation of coating properties during the test
	 experiment with the Ti trap.
	\item The experiment with a Ti trap proves that the part of the trap
	 surface uncoated by PFPE is negligible small ($4.4\times 10^{-7}$).
	\item Some flaking of PFPE coating in time would be discovered
	 immediately during the test experiment with the Ti trap.
\end{itemize}
Thus, according to the result of the test experiment with the Ti trap, we
have reasons to consider that the stability and repeatability of the PFPE
coating offer a much higher level of precision than the statistical
accuracy of measurements with a Be trap.

\subsection{Measurement data and extrapolation to the neutron lifetime}

\begin{figure}
\includegraphics[width=3.5in]{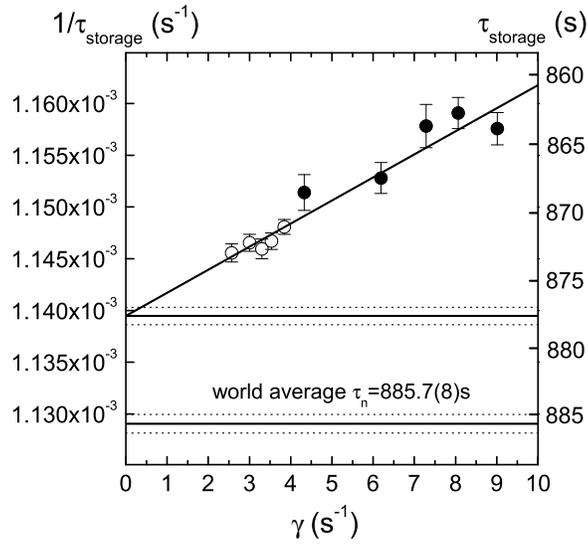}
\caption{\label{fig12} Result of extrapolation to the neutron
lifetime when combined energy and size extrapolations are used.
The open circles represent the results of measurements for a
quasispherical trap, and the full circles the results of
measurements for a cylindrical trap.}
\end{figure}

Figure~\ref{fig12} presents the results of measurements of the UCN
storage times for various energy intervals and different traps
(wide and narrow) as a function of the normalized loss rate
$\gamma $.
The UCN storage time was calculated according to
Eq.(\ref{eq19}). The values of $N_1$ and $N_2$ were corrected for
the background, which was measured at the end of each data-acquisition
cycle. The value of $N_1$ of the first emptying angle for the quasispherical
trap was corrected for the effect of UCN that have an energy higher than
the gravitational barrier. This correction is about 3.5 standard 
deviations ($3.5\sigma$) of this point. There are no other corrections for
$N_1$ and $N_2$. Corrections to the $\gamma$ function were discussed
in Sec.~\ref{ñalculation loss rate}. 

 Extrapolation of all the data to the neutron lifetime
yields a value of 877.60$\pm $0.65~s at $\chi ^2$=~0.95, which
means that a combined extrapolation is possible. However, if we
build an energy extrapolation for each trap and combine the two
results, we get 875.55$\pm $1.6~s.

To use the size extrapolation method, we have to combine the values
obtained from different traps within the same UCN energy range and
then calculate the average value of all the resultant values of
the neutron lifetime.

\begin{figure}
\includegraphics[width=3.5in]{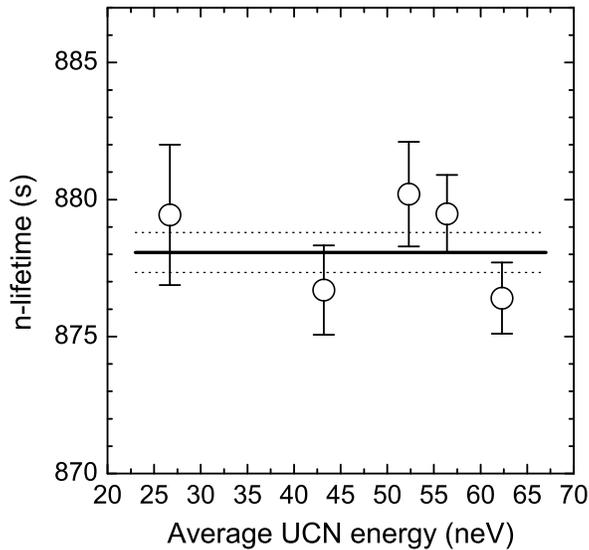}
\caption{\label{fig13} Extrapolated values of the neutron lifetime
for different mean UCN energies, when size extrapolation is used.
The solid, straight line corresponds to the average value of
neutron lifetime for given measurements.}
\end{figure}

Figure~\ref{fig13} demonstrates the results of size extrapolation
to the neutron lifetime for different energy ranges. The average
value of the neutron lifetime obtained by the size extrapolation
method was 878.07$\pm $0.73~s.

The results corresponding to these two methods differ by
1.5$\sigma $. The loss factor obtained in this experiment, $\eta =
2\times 10^{-6}$, agrees with the value found in the transmission
experiment \cite{Pok03}. For the final value of the neutron
lifetime we prefer using the result of size extrapolation, which
depends rather weakly on $\mu (E)$ and which we consider more
reliable.

\subsection{\label{Monte Carlo simulation}Monte Carlo simulation
 of the experiment and systematic errors}

To estimate the accuracy and test the reliability of the size
extrapolation method in which the value of the function $\gamma $
must be calculated, we performed a simulation of the experiment
using the Monte Carlo method.

The adopted Monte Carlo model describes the behavior of neutrons
with allowance for a gravitational field, the shape of the storage
traps, trap losses $\eta $ = 2$\times $10$^{-6}$, and the
geometries of the secondary volume and of the ultracold neutron
guide. The start spectrum for modeling is the spectrum measured in
the experiment. The angular distribution was prepared by simulation of
UCN storage in the trap for 50 s with 100{\%} diffuse reflections
without any losses before the start of the main simulation process. As
a result, we were able to simulate directly the measurement
procedure and build a time diagram for the counting rate at the
detector, similar to the one shown in Fig.~\ref{fig2}. The UCN 
storage time in the traps and the extrapolation to the neutron
lifetime that rely on the computed function $\gamma $ were
calculated in the same way as in the experiment. The single
adjustable parameter in the Monte Carlo model was the coefficient
of UCN diffuse scattering in the interaction with the trap
surface. All the information about the probability of mirror
reflection is extremely important. For instance, if it equals 
99.9{\%}, the behavior of UCN in the trap becomes strongly
correlated and the resulting prediction becomes extremely 
difficult to make.

\begin{figure}
\includegraphics[width=3.5in]{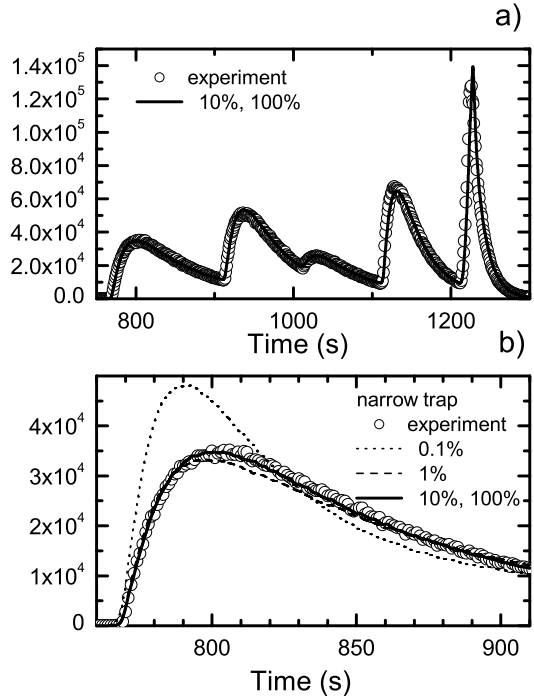}
\caption{\label{fig14} Simulation of an experiment by the Monte
Carlo method, consisting in simulating the neutron discharge from
a narrow cylindrical trap. The dotted curve corresponds to the
results of calculations with a 0.1{\%} diffuse reflection
probability; the dashed curve corresponds to a 1{\%} diffuse
reflection probability, and the solid curve to 10{\%} and
100{\%}.}
\end{figure}

A comparison of the results of Monte Carlo calculations for
different values of the diffuse scattering probability and the
experimental results (Fig.~\ref{fig14}) makes it possible to
conclude that the probability of UCN diffuse scattering by the
PFPE coating amounts to about 10{\%}. In Fig.~\ref{fig14}(a) we
compare the experimental diagram and the Monte Carlo simulation
diagram, obtained with the diffuse scattering coefficients
equal to 10{\%} and 100{\%}. The experiment is successfully
described for both diffusivity values. However, when the diffuse
scattering probability is 0.1{\%}, the agreement between the
calculated and experimental results becomes unsatisfactory. The
results of such calculations for the first part of the time
diagram, which is the most sensitive to the neutron mirror reflection, are
shown on a larger scale in Fig.~\ref{fig14}(b).

\begin{figure}
\includegraphics[width=3.5in]{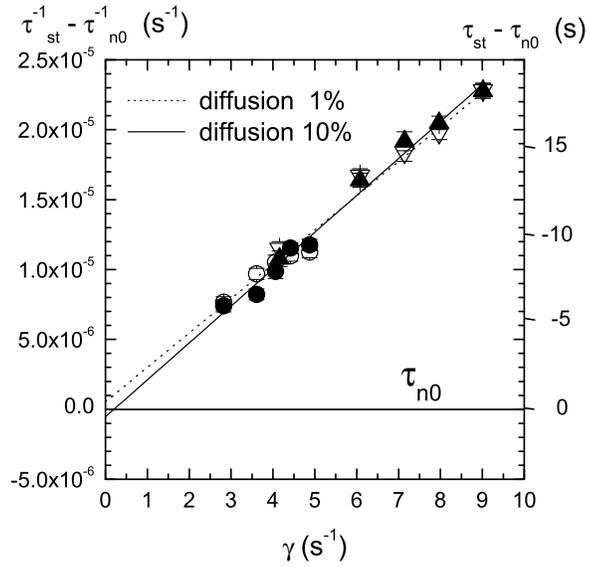}
\caption{\label{fig15} Monte Carlo experiment with a 1{\%} and
10{\%} diffuse reflection probability involving simulation of an
extrapolation to the neutron lifetime. The circles represent the
results of simulation for a wide trap and triangles the results of
simulation for a narrow cylindrical trap. The open and full
figures correspond to 1{\%} and 10{\%} diffuse reflection
probability respectively.}
\end{figure}

The final simulation of the experiment was made for 10{\%} and
1{\%} diffuse reflection probabilities. The model storage times
extrapolated to the neutron lifetime for the wide and narrow
cylindrical traps and five different UCN energy ranges are
presented in Fig.~\ref{fig15}. To simplify the Monte Carlo
calculations for the wide trap, we used cylindrical traps instead
of quasispherical ones. In the final analysis of the data obtained with
this model, we reproduced the value of the neutron lifetime
adopted in the calculation with an accuracy of $\pm $0.236~s. This
accuracy was limited by the statistical accuracy of the Monte
Carlo calculations. About 10$^{9}$ neutrons were used in the
simulations presented in Fig.~\ref{fig15}. This number
corresponds to the beginning of the monitoring process. Thus, because
we employed the computed value of the function $\gamma $, the
systematic uncertainty of the size extrapolation method amounted
to $\pm $0.236~s.

\begin{figure}
\includegraphics[width=3.5in]{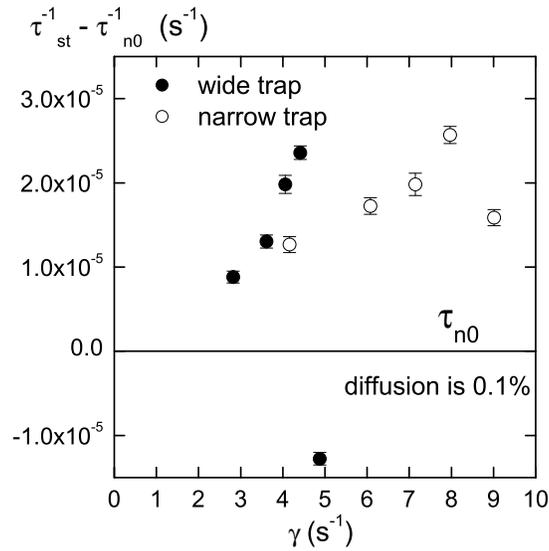}
\caption{\label{fig16} Monte Carlo experiment with a 0.1{\%}
diffuse reflection probability involving simulation of an
extrapolation to the neutron lifetime. The full circles represent
the results of simulation for a wide trap and the open circles
the results of simulation for a narrow cylindrical trap.}
\end{figure}

It should be mentioned that Monte Carlo simulations with a probability
of diffusive scattering of 0.1{\%} have also been carried out.
Figure~\ref{fig16} demonstrates the result of this simulation.
All points besides the first emptying show the shorter storage time
because the cleaning of spectrum becomes less effective and neutrons
leave the trap even during the storage period.
The first emptying should be excluded because it is near the edge of
the energy spectrum in the Monte Carlo calculations. This picture 
visibly disagrees with the experimental plot (see Fig.~\ref{fig12}),
leading us to conclude that high mirror reflection did not take place
in the experiment.

\begin{figure}
\includegraphics[width=3.5in]{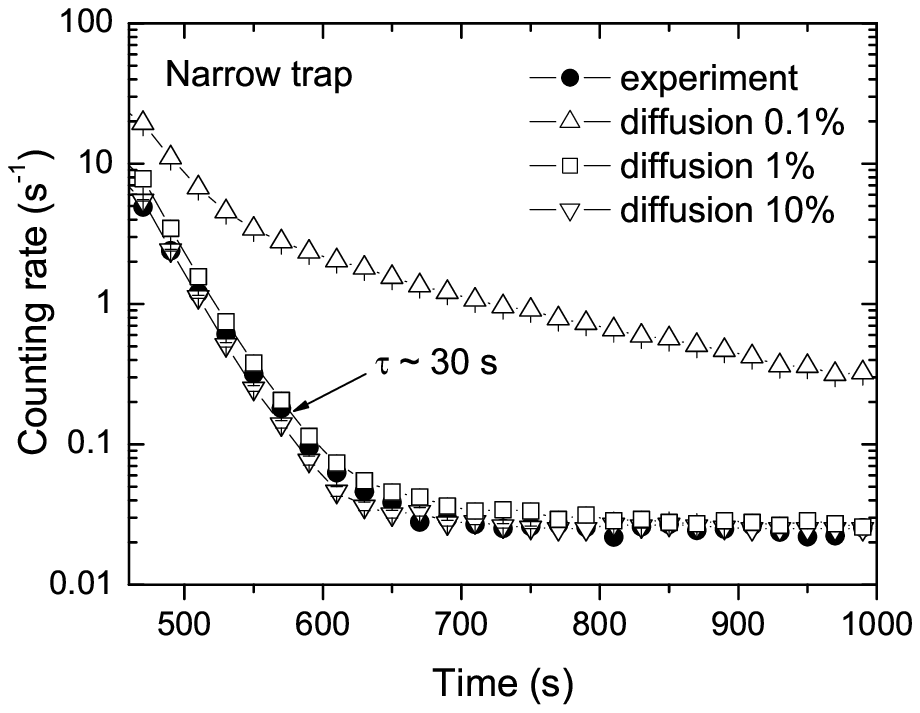}
\caption{\label{fig17} Monte Carlo simulation of leakage process
of UCN exceeding the gravitational barrier of the trap from the
narrow cylindrical trap. The results of Monte Carlo calculations with
different diffuse reflection probability are presented: by up
triangles with 0.1{\%} diffusion, by squares with 1{\%} diffusion
and by down triangles with 10{\%} diffusion. The filled circles
represent the results of experiment.}
\end{figure}

\begin{figure}
\includegraphics[width=3.5in]{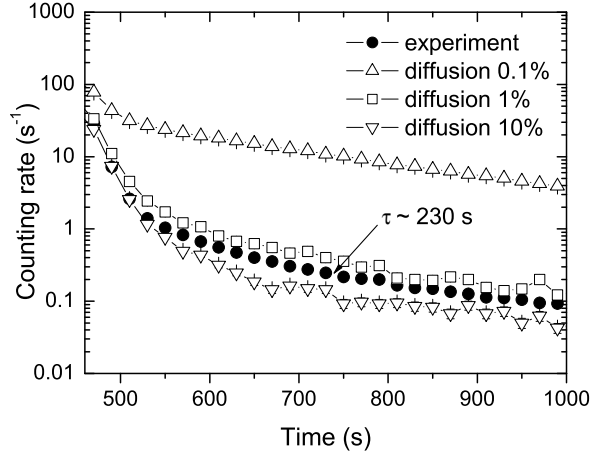}
\caption{\label{fig18} Monte Carlo simulation of leakage process
of UCN exceeding the gravitational barrier of the trap from the
quasispherical trap. The results of Monte Carlo calculations with different
diffuse reflection probability are presented: by up triangles with
0.1{\%} diffusion, by squares with 1{\%} diffusion and by down
triangles with 10{\%} diffusion. The filled circles represent the
results of experiment.}
\end{figure}

Another piece of evidence for the absence of high mirror reflection
comes from the analysis of the leakage process of UCN exceeding the
gravitational barrier of the trap just after the beginning of storage
process, which is shown in Fig.~\ref{fig17} for a narrow
cylindrical trap and in Fig.~\ref{fig18} for a quasispherical
trap. Monte Carlo simulation of this process shows that with
high mirror reflection the tail of the time spectra of leaked
neutrons has to be very long. The results of the experiment are
consistent with the results of Monte Carlo simulation for diffuse
reflection probability of more than 1{\%}.

\subsection{\label{residual gas section}Effect of residual gas
 on UCN storage}

When the neutron lifetime is measured with a high precision, the
effect of residual vacuum on the UCN loss cannot be ignored. For
instance, a residual gas pressure of $5\times 10^{-6}$ mbar would
introduce an error of about 0.4~s into the value of the UCN
lifetime in the trap, which is comparable to the statistical
accuracy of the measurements. Such a correction cannot be measured
directly (e.g., by improving a vacuum pressure by an order of
magnitude) because the expected effect is smaller than the
statistical uncertainty.

\begin{figure}
\includegraphics[width=3.5in]{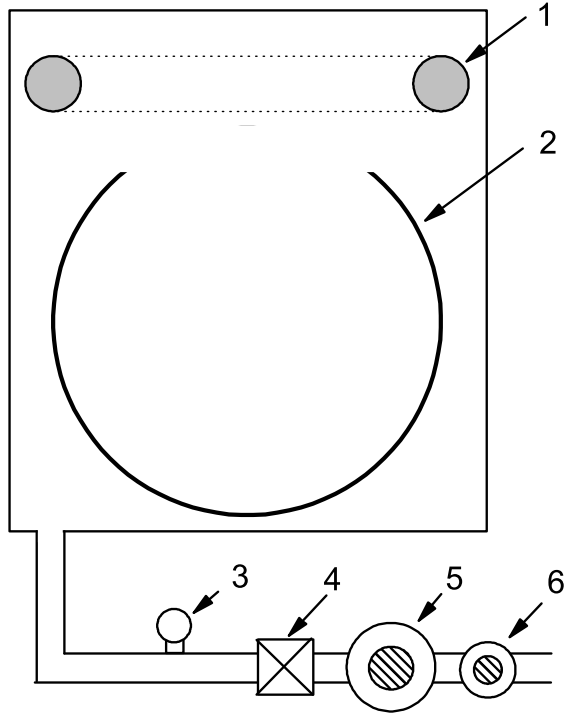}
\caption{\label{fig19} Vacuum system of ``Gravitrap''
installation: 1 - cryogenic vacuum pump; 2 - UCN trap; 3 - vacuum
gauge; 4 - vacuum valve; 5 - turbo pump {\#}1; 6 - turbo pump
{\#}2.}
\end{figure}

The vacuum system of an installation is shown in Fig.~\ref{fig19}. A
``clean'' vacuum volume is connected with turbo pump {\#}1 by
a pipe about 1.2 m of length and 90 mm in diameter. Turbo pump
{\#}1 has a pumping speed of about 550 l/s. Turbo pump {\#}2
with a smaller pumping speed is connected to the output of the
first turbo pump. A vacuum valve with a pneumatic drive allows
separation of the main volume from the pumping system.

As one can learn from Fig.~\ref{fig19} the vacuum gauge is
placed rather close to the turbo pump. There exists a pressure gradient  
in the vacuum pipe when the vacuum valve is open and turbo pump is on.
In this situation, the vacuum gauge will detect some value of
vacuum that can be far from the equilibrium value of vacuum for
our system. Detecting the equilibrium value of vacuum requires some
special actions. The vacuum valve was closed for some
time. During several seconds the gradient of vacuum owing to the turbo
pump disappears and we took the equilibrium value of vacuum from
the vacuum gauge. All values of vacuum mentioned in the following 
were measured according to this simple procedure.

Of course, this procedure of measurement of equilibrium vacuum is
valid only for a vacuum system without any temperature
gradients. In our system, the main vacuum volume has a
temperature of about --160$^\circ$C, whereas the vacuum gauge is
at room temperature. In this case, the ``real'' vacuum in the main
vacuum volume ${P}'$ will differ from the vacuum detected with
the vacuum gauge, $P$, by a temperature-dependent factor $k$: ${P}'=k(T_1
,T_2 )\cdot P$, where $T_1 $ and $T_2 $ are the temperatures of
the main vacuum volume and vacuum pipe. The factor $k$ can be
calculated with some precision, but it is important for us that
this factor does not depend on the pressure, because we are going
to extrapolate to zero vacuum using the measurements at the
two different vacuums.

The main measurements of the UCN storage time were carried out at
a vacuum pressure of $5\times 10^{-6}$ mbar (equilibrium value).
The result of our size extrapolation to the neutron lifetime
(878.1~s) differs from the world average value (885.7~s) by 7.6~s,
or $6.5\sigma$. At a vacuum pressure of $5\times 10^{-6}$ mbar the
correction for neutron lifetime can be about 1~s. Nevertheless,
we decided to improve the vacuum in our storage volume. Using
a turbo pump with a higher pumping speed makes no sense, because
the pumping speed of the existing turbo pump (550 l/s) was already
limited by the conductivity of the vacuum pipe. The best decision
was to install a cryogenic vacuum pump directly inside the main
vacuum volume, very close to the UCN trap. The position of the 
cryogenic pump is shown in Fig.~\ref{fig19}.
It is a torus made of a 100-mm-diameter copper pipe . The
surface of the cryopump is about 1 m$^{2}$. To ``switch on'' the
cryopump one has to fill it with liquid helium.

After a liquid-helium cryogenic pump was mounted near the storage
volume, the vacuum measured on the gauge was improved from
$5\times 10^{-6}$ to $3\times 10^{-7}$ mbar. Even closing
the vacuum valve before the turbo pump does not change its
value. This means that the vacuum near the UCN trap is at least not
worse than $3\times 10^{-7}$ mbar.

The UCN storage time was measured with improved vacuum. The
results of measurements for quasispherical and narrow traps are
presented in Fig.~\ref{fig11}. The best accuracy was achieved for
a quasispherical trap: the UCN storage time is 872.36$\pm $0.56~s
if the cryopump is on, and is 872.2$\pm $0.3~s (an average storage
time for all six runs) if the cryopump is off. Therefore, one order
of magnitude improvement of vacuum rises the UCN storage time by
0.2$\pm $0.6~s. We can conclude that a vacuum of $5\times 10^{-6}$
mbar could decrease the UCN storage time by less than 1~s (90{\%} 
confidence limit).

So, we can not explain the discrepancy between the measured and the world
average neutron lifetime values by poor vacuum. Moreover,
the uncertainty of these measurements with the improved vacuum
($\pm $0.6~s) is comparable to the statistical accuracy of our neutron
lifetime result. It would be nice to measure the correction for
vacuum more precisely.

Measuring the correction for vacuum more precisely can be done by
extrapolating to zero vacuum by increasing the residual gas pressure.
The only condition is that we save the same residual gas
content. To save it, we reduced the pumping speed of our system.
In practice, this means that we switched off turbo pump {\#}1
(see Fig.~\ref{fig19}) and switched on turbo pump {\#}2, which has
a lower pumping speed. In this way we increased the residual gas 
pressure to $8\times 10^{-4}$ mbar and measured the UCN storage
time at this vacuum. The main measurements of the UCN storage time
were carried out at a vacuum of $5\times 10^{-6}$ mbar. Then, the
inverse UCN storage times as a function of vacuum were extrapolated
to zero vacuum. The difference between the UCN storage time at zero
vacuum (extrapolated) and the UCN storage time at a vacuum of
$5\times 10^{-6}$ mbar is the vacuum correction, which amounted to
0.4$\pm $0.02~s. This correction does not depend on the UCN energy
and can be applied to refine the result for the neutron lifetime.

The correction $\Delta \tau _{vac}^{ - 1}$ for UCN losses from vacuum was
calculated according to the equation:
\begin{equation}
\label{eq25}
\Delta \tau _{vac}^{ - 1} =\frac{\tau _{vac2}^{ - 1} -\tau _{vac1}^{ - 1}}
{P_2 -P_1 }\cdot P_1
\end{equation}
where $\tau _{vac2}^{ - 1}$ and $\tau _{vac1}^{ - 1}$ are inverse
UCN storage times at vacuum pressures of $P_2 =8\times 10^{-4}$ and $P_1 =5\times
10^{-6}$ mbar, respectively. It is important to note that the
value of correction depends only on the ratio of $P_2 $ and $P_1 $.
This means that the temperature-dependent factor $k(T_1 ,T_2 )$, which
we discussed earlier, will cancel in Eq.(\ref{eq25}). So we can use
the values of vacuum that were measured in a ``warm'' position of
the vacuum gauge. This procedure is correct and is all we
need for the neutron lifetime measurement. We do not need to use any other
information about the rest gas content. For us it is an unknown gas.

Nevertheless, one can estimate the parameter $p\tau$ for this gas.
In case of a vacuum equilibrium state we can use the condition that
the flux of molecules in a warm part of the system, $\phi _{warm} =\rho
_{warm} \cdot v_{warm}$, is equal to the flux of molecules in a cold
part of the system, $\phi _{cold} =\rho _{cold} \cdot v_{cold} $. 
Thus the loss factor for UCN in a cold part of the installation is the
same as in a warm part of the installation (and although the final energy of
upscattered UCN is different in a warm part and in a cold part, in
any case, UCN will be lost). This approach made it possible to
calculate the parameter $p\tau$ for the residual gas (9.5
mbar$\cdot $s) and draw some conclusions about the rest gas
content. We can compare this $p\tau$ value with ones measured for
different gases. It is very probable that the rest vacuum contains
mainly the air and about one-fourth of the hydrogen. This conclusion
is not important for our calculation of vacuum correction and
gives us only an estimate of how much hydrogen is in our system.

In conclusion, the method used for the calculation of correction of UCN
losses owing to the rest vacuum does not need a direct measurement of
vacuum in a cold part of the installation nor does it require a residual gas
analysis.

\subsection{Final result for the neutron lifetime and a list
 of systematic corrections and errors }
 
The magnitudes of the systematic effects and their uncertainties
are listed in Table~\ref{table}.

\begin{table*}
\caption{\label{table}Systematic effects and their uncertainties.}
\begin{ruledtabular}
\begin{tabular}{lcr}
Systematic effect& Magnitude (s)& Uncertainty (s) \\
\hline
Method of calculating $\gamma $& 0& 0.236 \\
Influence of shape of function $\mu $(E) & 0& 0.144 \\
UCN spectrum uncertainty& 0& 0.104 \\
Uncertainty of trap dimensions (1 mm)& 0& 0.058 \\
Residual gas effect& 0.4& 0.024 \\
Uncertainty in PFPE critical energy (20 neV)& 0& 0.004 \\
\hline
Total systematic correction& 0.4& 0.3 \\
\end{tabular}
\end{ruledtabular}
\end{table*}

The main contribution to the uncertainty is provided by the
statistical accuracy of determining the UCN lifetime. The next most 
important (by value) uncertainty is that in the calculation of the function
$\gamma $. The contributions from the uncertainty of the shape of
the function $\mu(E)$ and the uncertainty in the UCN spectrum, which are
much smaller, were estimated by varying their parameters within
the uncertainty limits allowed by the experimental data. Thus, the
total systematic correction proved to be equal to 0.4~$\pm
$~0.3~s, and the final result for the neutron lifetime is
$878.5\pm0.7_{stat}\pm 0.3_{sys}$ s.

\section{Reasons for the discrepancy between the results
 of experiments on UCN storage }
 
The new result for the neutron lifetime differs from the world
average value by 6.5$\sigma $, although actually this deviation is
determined mainly by the discrepancy between our result and the
result of the experiment done by Arzumanov \textit{et al.} \cite{Arz00},
who also achieved a high precision in their measurements. The
discrepancy with the work of Arzumanov \textit{et al.}\cite{Arz00} is
5.3$\sigma $. It is
extremely difficult to discuss the reasons for this discrepancy.
We are sure about the results of our experiment, since the
probability of UCN losses amounts only to 1{\%} of the probability
of neutron $\beta $ decay, whereas in the experiment by Arzumanov 
\textit{et al.} \cite{Arz00} the probability of UCN losses amounts to about
30{\%}.

In our experiment we have an extrapolation to a neutron lifetime that is
only 5~s from the best experimental storage time. In experiment 
of Arzumanov \textit{et al.}\cite{Arz00} the task was to extrapolate to
about 120 s. It is not completely clear
how to reach a systematic accuracy of the extrapolation of 0.4 s
by using two experimental storage times (approximately 765 and 555
s) for two configurations of the UCN trap with a different neutron free
path.

In our experiment, we used solid PFPE, and the process of neutron
quasielastic scattering was completely suppressed. Unfortunately,
the experiment by Arzumanov \textit{et al.} \cite{Arz00} was carried out
before the effect of quasielastic scattering by liquid Fomblin was
discovered, and the authors of Ref. \cite{Arz00} have to analyze
the effect of quasielastic scattering on experimental results.

We are forced to note that, in all the experiments involving liquid
Fomblin \cite{Arz00,Mam93,Mam89}, neutron quasielastic scattering
was revealed and the process was found to change the spectrum
during UCN storage. Although these experiments were run in a
scaling mode (i.e., for traps of different dimensions the neutron
containment time was chosen in such a way that the number of
collisions was the same), the effect of spectrum change caused by
quasielastic scattering was not taken into account. Our
preliminary analysis shows that this effect may lead to an
overestimated value of the extrapolated neutron lifetime.

We performad a Monte Carlo simulation of experiment \cite{Mam89}. These
calculations were reported \cite{Fom05,Fom07} at the 5th and 6th
International UCN Workshops ``Ultracold and Cold Neutrons. Physics
and Sources.''. In Refs.\cite{Fom05,Fom07} we showed that
the correction for uncorrected neutron lifetime from the experiment
\cite{Mam89} is 2.5~s, instead of 9~s, which was used in Ref.
\cite{Mam89}. Therefore, the new corrected value from the experiment
of Ref.\cite{Mam89} will be 881.5$\pm $3.0~s, which is in the agreement
with our result \cite{Ser05} of 878.5$\pm $0.8~s. Additionally, it
should be mentioned that the effect of quasielastic scattering is the
most significant when the initial UCN spectrum in the experiment has an
upper cutoff higher than the critical energy of Fomblin. In this
case there is regular leakage of UCN through the energy barrier from
the first moment of the storage process. In the case when the initial spectrum
is cut below the critical energy of Fomblin, the effect of UCN leakage
through the energy barrier appears only at long holding times.
Therefore, the correction is considerably lower. Indeed, the experiment
MAMBO~II \cite{Pic00} used a UCN spectrum with a cutoff below the Fomblin
critical energy and a neutron lifetime value of 881$\pm $3~s has been
obtained. This result is in reasonable agreement with our result
\cite{Ser05} of 878.5$\pm $0.8~s as well as with the corrected value of
the experiment in Ref.\cite{Mam89} 881.5$\pm $3.0~s. Thus, the main reasons for
the discrepancy are beginning to become clear. They are connected with
quasielastic scattering on the liquid Fomblin and use of
an initial UCN spectrum with a cutoff higher than the critical Fomblin
energy. A detailed article devoted to Monte Carlo simulation of
the experiment in Ref.\cite{Mam89} is in progress.

Finally, we have to discuss the discrepancy with our previous result
obtained with a solid oxygen coating of the trap \cite{Nes92}.
This discrepancy is not so big and is 2.7$\sigma $. Therefore, we cannot
analyze the reason for this discrepancy; we can only discuss the
possible assumptions. We cannot exclude the possible 
contribution of anomalous losses with a different energy dependence.
Although the dependence of the final result on $\mu(E)$ is
considerably suppressed when the size extrapolation is used, in our
previous experiment we used a combined extrapolation in which the
effect could have been suppressed to a lesser extent.
Unfortunately, the level of statistical accuracy of the previous
experiment makes an analysis of these assumptions impossible.
Furthermore, it must be noted that the coating properties of solid
oxygen are inferior to those of PFPE. We can assume that the quality
of coating by solid oxygen was a little bit worse for the narrow trap,
because the trap configuration make it more difficult to achieve
uniform coating. The portion of the uncoated surface for solid oxygen
was approximately $10^{-2}$, whereas that for PFPE awas (4.4~$\pm $~1.3)$\times
$10$^{-7}$. The most important problem of the equivalence of the
coatings for the wide and narrow traps has been solved more
reliably in our recent experiment that uses PFPE.
Thus, our experiment with PFPE possessed a very small loss factor.
It also exhibited no effects of anomalous losses and neutron quasielastic
scattering, in contrast to other experiments. All this guarantees
the reliability of our results.

\section{Conclusion}
In conclusion we would like to summarize the main advantages of our
experiment:
\begin{enumerate}
\item The storage time in the experiment is the closest to
a neutron lifetime. The probability of losses is about 1{\%} of the
probability of a neutron $\beta$ decay. 
\item The extrapolated time differs from the best storage time by only
5 s, whereas the accuracy of extrapolation is $\pm $0.7$_{stat}$ s and
$\pm $0.3$_{sys}$ s. This means that the relative accuracy of taking
into account the losses in the storage process is only about 10{\%}.
\item The process of quasielastic scattering is completely suppressed.
The upper limit for corrections of such processes is 0.03~s.
\item The coating properties of PFPE are nearly ideal. The uncovered part of
the surface has to be less than 10$^{-6}$, guaranteeing 
the same loss factor for the different traps with beryllium
substrate.
\item The stability and reproducibility of PFPE coating
were demonstrated in the course of the experiment.
\end{enumerate}
All these advantages of this experiment allow us to obtain the most
precise result of the neutron lifetime measurement: 878.5$\pm $0.8~s.

\begin{acknowledgments}
The authors are grateful to V.Alfimenkov, V.Lushchikov,
A.Strelkov and V.Shvetsov for their contribution at the initial
stage of the development of the installation; A.Steyerl, O.Kwon,
and N.Achiwa for their participation in measurements and fruitful
discussions; T.Brenner for intensive and very helpful
assistance during the experiment; PSI for help in
manufacturing UCN traps; the Russian Foundation of Basic Research
for support under Contract No. 02-02-17120; and the Russian Academy of
Sciences program ``Physics of Elementary Particles''.
\end{acknowledgments}

\end{document}